\documentclass[prl,aps,twocolumn,showpacs]{revtex4} 
\usepackage{graphicx} 
\usepackage{dcolumn} 
 
\graphicspath{{fig/}} 

\begin{document} 
 
\title{ 
Inherent spin density wave instability by vortices in superconductors
with strong Pauli effects 
} 

\author{K. M. Suzuki} 
\affiliation{Department of Physics, Okayama University, 
Okayama 700-8530, Japan} 
\author{M.  Ichioka} 
\affiliation{Department of Physics, Okayama University, 
Okayama 700-8530, Japan} 
\author{K. Machida} 
\affiliation{Department of Physics, Okayama University, 
Okayama 700-8530, Japan} 
\date{\today}

\begin{abstract} 
A novel spin density wave (SDW) instability mechanism enhanced by vortices under fields
is proposed to explain the high field and low temperature (HL) phase in 
CeCoIn$_5$.
In the vortex state the strong Pauli effect and the nodal
gap conspire to enhance the momentum resolved 
spectral weight exclusively along the nodal direction over the normal value,
providing a favorable nesting condition for SDW with ${\bf Q}=(2k_F, 2k_F, 0.5)$ 
only under high field ($H$).
Observed mysteries of the field-induced SDW confined within $H_{c2}$
are understood consistently, such facts that  ${\bf Q}$ is directed to the nodal direction independent of $H$,
SDW diminishes under tilting field from the $ab$ plane,
and the SDW transition line in $(H,T)$ has a positive slope.
\end{abstract} 
 
 
%74.70.Tx	Heavy-fermion superconductors 
%74.25.Uv	Vortex phases (includes vortex lattices, vortex liquids, and vortex glasses)
%74.20.-z Theories and models of superconducting state   
%74.20.Rp Pairing symmetries (other than s-wave)   
%74.25.-q Properties of type I and type II superconductors   
%74.25.Qt Vortex lattices, flux pinning, flux creep   

\maketitle 
%%%%%%%%%%%%%%%%%%%%%%%%%%%%%%%%%%%%%%%%%%%%%%%%%%%%%%%%%%%%%%%% 
%\section{introduction} 
The competing-order phenomena are a hallmark of strongly correlated 
systems. This is particular true for superconductors, such as in heavy Fermion
materials, high T$_c$ cuprates or pnictides
since competing magnetism is deeply rooted to the pairing mechanism itself
where an SDW  order generically competes
or coexists with superconductivity (SC)\cite{chris}.
Thus  it is not entirely surprising to see that applied field induces an SDW in a superconductor.
In fact, in La$_{1-x}$Sr$_x$CuO$_4$\cite{mesot} and CeRhIn$_5$
under pressure\cite{flouquet} the field induced SDW is observed only for finite fields
and is absent at zero field.
A remarkable observation in CeCoIn$_5$ is that the induced SDW is confined 
exclusively in the superconducting state below the upper critical field $H_{c2}$\cite{chris}
(see HL in Fig.3(c)).

Since Abrikosov\cite{abrikosov}, there have been many studies on vortices
 in a type II superconductor based on a concept that the vortex core is a 
 featureless rigid cylindrical object filled with the normal electrons.
An emergent new concept based on microscopic Bogoliubov-de Gennes or
quasi-classical Eilenberger framework is recently revealing a far richer quasi-particle (QP) 
structure both in real space and reciprocal space which governs 
a variety of physical characteristics in a type II superconductor
under a field\cite{hasegawa}.
In particular, in a superconductor characterized by an anisotropic
gap, including a nodal gap such as a d-wave symmetry, the QP spatial
structure is directly or indirectly measured by various experimental methods, such as
scanning tunneling microscope (STM) as direct spatial images\cite{hess}, 
small angle neutron scattering (SANS)
as Fourier-transformed images\cite{bianchi}, or field-angle resolved specific heat\cite{an} 
or thermal conductivity\cite{izawa}.
With this emergent QP concept we would expect to uncover new phenomena.

Here we investigate detailed behaviors of QPs induced by vortices 
with the nodal gap structure (i.e. $d_{x^2-y^2}$)
and the Pauli paramagnetic effect to uncover the origin of the HL phase
in a heavy fermion superconductor CeCoIn$_5$.
This has been discussed theoretically from different view points\cite{littlewood,tsunetsugu,yanase,miyake,ikeda}.
We find generic features of the QP behavior, which 
enables us to draw vivid physical picture, synthesizing the real space and reciprocal space
complemental information.
This picture, in particular  in reciprocal space, leads naturally us to an SDW instability
in a high field, which is confined in the superconducting state.
The duality of the QP behaviors in real and reciprocal space is
a key concept to uncover the origin of the HL phase. Note in passing that in dHvA effect
in the superconducting state the QPs executing the cyclotron motion in real space carry information in the 
original Fermi surface topology in the normal state, evidencing the dual nature of the QPs created 
around the vortex core.

The mysterious HL phase exists for both $H$ applied to the basal plane 
($H$$\parallel$$ab$)\cite{losalamos,kakuyanagi,vesna,curro,matsuda}
and the $c$-axis ($H$$\parallel$$c$)\cite{kumagai} of the tetragonal crystal. 
In particular, for $H$$\parallel$$ab$ case
the mounting evidence\cite{michel1,michel2} shows that an SDW characterizes the HL phase. 
The HL phase (see Fig.3(c)) appears via a second order phase transition $H_{Q}(T)$ from the 
conventional Abrikosov vortex lattice state.
Namely the sublattice magnetization of SDW grows continuously at
$H_{Q}(T)$ and disappears abruptly at $H_{c2}$ via a first order transition.
The ordering wave vector $Q=(0.45, 0.45, 0.5)$ in the reciprocal vector units
is independent of the field directions $H$$\parallel$$(1\bar{1}0)$\cite{michel1} 
and $H$$\parallel$$(100)$\cite{michel2},
and independent of the field strength.
The phase boundary $H_{Q}(T)$ between the HL and the Abrikosov state is
 almost independent of two field directions. $H_{Q}(T)$ is an increasing function of $T$.
Those features mainly come from the neutron scattering experiments\cite{michel1,michel2} and basically
are consistent with other thermodynamic measurements\cite{losalamos} and NMR experiments\cite{vesna,curro}.
Since the pairing symmetry of this material is firmly established as $d_{x^2-y^2}$ type\cite{an,izawa},
the origin of this HL phase must be tied to (1) the $d$-wave nature of this system and (2)
the vortex state under a field.
As for  (1) the controversy of the pairing symmetry either $d_{x^2-y^2}$
or $d_{xy}$ is now resolved\cite{izawa,aoki} and there is little doubt for the $d_{x^2-y^2}$ symmetry in 
CeCoIn$_5$\cite{an}.
As for (2) it is known by SANS experiments\cite{bianchi} that the Pauli paramagnetic
effect is indispensable in understanding the anomalous behaviors of the scattering
form factor as functions of field strength and temperature.
The purpose of this paper is to examine the generic SDW instability enhanced by the presence of vortices
under the strong paramagnetic effect  on the basis of the full selfconsistent solutions of 
microscopic quasi-classical Eilenberger equations.

%%%%%%%%%%%%%%%%%%%%%%%%%%%%%%%%%%%%%%%%%%%%%%%%%%%%%%%%%%%%%%%% 
%\section{formulation and notations} 

We calculate the spatial structure of the vortex lattice state 
by quasiclassical Eilenberger theory in the clean 
limit valid for $k_{\rm F}\xi \gg 1$ ($k_{\rm F}$ is the Fermi wave number and 
$\xi$ is the superconducting coherence length)~\cite{hasegawa}.
The Pauli paramagnetic effects are included through the Zeeman term $\mu_{\rm B}B({\bf r})$, 
where $B({\bf r})$ is the flux density of the internal field and 
$\mu_{\rm B}$ is a renormalized Bohr 
magneton.
The quasiclassical Green's functions
$g( \omega_n +{\rm i} {\mu} B, {\bf k},{\bf r})$, 
$f( \omega_n +{\rm i} {\mu} B, {\bf k},{\bf r})$, and 
$f^\dagger( \omega_n +{\rm i} {\mu} B, {\bf k},{\bf r})$  
are calculated in the vortex lattice state  
by the Eilenberger equations 
\begin{eqnarray} &&
\left\{ \omega_n +{\rm i}{\mu}B 
+\tilde{\bf v} \cdot\left(\nabla+{\rm i}{\bf A} \right)\right\} f
=\Delta\phi g, 
\nonumber 
%\label{eq:eil1}
\\ && 
\left\{ \omega_n +{\rm i}{\mu}B 
-\tilde{\bf v} \cdot\left( \nabla-{\rm i}{\bf A} \right)\right\} f^\dagger
=\Delta^\ast \phi^\ast g  , \quad 
%\label{eq:eil2}
\label{eq:Eil}
\end{eqnarray} 
where $g=(1-ff^\dagger)^{1/2}$, ${\rm Re} g > 0$, 
$\tilde{\bf v}={\bf v}/v_{{\rm F}0}$, 
and the Maki parameter ${\mu}=\mu_{\rm B} B_0/\pi k_{\rm B}T_{\rm c}$. 
${\bf k}=(k_a,k_b,k_c)$  is the relative momentum of the Cooper pair, 
and ${\bf r}$ is the center-of-mass coordinate of the pair. 
We set the pairing function 
$\phi({\bf k})=\sqrt{\mathstrut 2}(k_a^2-k_b^2)/(k_a^2+k_b^2)$ in $d_{x^2-y^2}$-wave pairing.
We use the Eilenberger units ${\bf R_0}$ and $B_0$~\cite{ichioka}.
The energy $E$, pair potential $\Delta$ and Matsubara frequency $\omega_l$ 
are in units of $\pi k_{\rm B} T_{\rm c}$. 

The averaged Fermi velocity on the Fermi surface 
$v_{\rm F0}=\langle v^2 \rangle_{\bf k}^{1/2}$ 
where $\langle \cdots \rangle_{\bf k}$ indicates the Fermi surface average. 
To model the quasi-two dimensional Fermi surface of ${\rm CeCoIn_5}$
we use a Fermi surface with warped cylinder-shape coming from
the so-called $\alpha$-orbit (see four cylinders in Fig.3(a))~\cite{harima}
and the Fermi velocity is given by 
${\bf v}=(v_a,v_b,v_c) \propto 
(\cos\theta_k,\sin\theta_k, \tilde{v}_z \sin k_c)$
at the Fermi surface 
${\bf k}\propto(k_{\rm F0}\cos\theta_k,k_{\rm F0}\sin\theta_k,k_c)$ 
with $-\pi \le \theta_k \le \pi$ and $-\pi \le k_c \le \pi$.
We set $\tilde{v}_z =0.5$, thus the anisotropy ratio 
$\gamma={\xi_c}/{\xi_{ab}} \sim  
{\langle v_c^2 \rangle_{\bf k}^{1/2}}/{\langle v_a^2 \rangle_{\bf k}^{1/2}} 
\sim 0.5$
as observed in ${\rm CeCoIn_5}$.

As for selfconsistent conditions, 
the pair potential is calculated by 
\begin{eqnarray}
\Delta({\bf r})
= g_0N_0 T \sum_{0 < \omega_n \le \omega_{\rm cut}} 
 \left\langle \phi^\ast({\bf k}) \left( 
    f +{f^\dagger}^\ast \right) \right\rangle_{\bf k} 
\label{eq:scD} 
\end{eqnarray} 
with 
$(g_0N_0)^{-1}=  \ln T +2 T
        \sum_{0 < \omega_n \le \omega_{\rm cut}}\omega_l^{-1} $. 
We use $\omega_{\rm cut}=20 k_{\rm B}T_{\rm c}$.
The vector potential for the internal magnetic field 
is selfconsistently determined by 
\begin{eqnarray}
\nabla\times \left( \nabla \times {\bf A} \right) 
=\nabla\times {\bf M}_{\rm para}({\bf r})
-\frac{2T}{{{\kappa}}^2}  \sum_{0 < \omega_n} 
 \left\langle \tilde{{\bf v}}
         {\rm Im} g  
 \right\rangle_{\bf k}, 
\label{eq:scH} 
\end{eqnarray} 
where we consider both the diamagnetic contribution of 
supercurrent in the last term 
and the contribution of the paramagnetic moment 
${\bf M}_{\rm para}({\bf r})=(0,0,M_{\rm para}({\bf r}))$ 
with 
\begin{eqnarray}
M_{\rm para}({\bf r})
=M_0 \left( 
\frac{B({\bf r})}{H} 
- \frac{2T}{{\mu} H }  
\sum_{0 < \omega_n}  \left\langle {\rm Im} g 
 \right\rangle_{\bf k}
\right) . 
\label{eq:scM} 
\end{eqnarray} 
The normal state paramagnetic moment 
$M_0 = ({{\mu}}/{{\kappa}})^2 H $,   
${\kappa}=B_0/\pi k_{\rm B}T_{\rm c}\sqrt{8\pi N_0}$  and 
$N_0$ is the density of states (DOS) at the Fermi energy in the normal state. 
We set the Ginzburg-Landau parameter $\kappa=89$.
We solve eq. (\ref{eq:Eil}) and eqs. (\ref{eq:scD})-(\ref{eq:scM})
alternately, and obtain selfconsistent solutions
as in previous works~\cite{hasegawa}
under a given unit cell of the vortex lattice.

%%%%%%%%%%%%%%%%%%%%%%%%%%%%%%%%%%%%%%%% 
\begin{figure}[tb] 
\includegraphics[width=7.2cm]{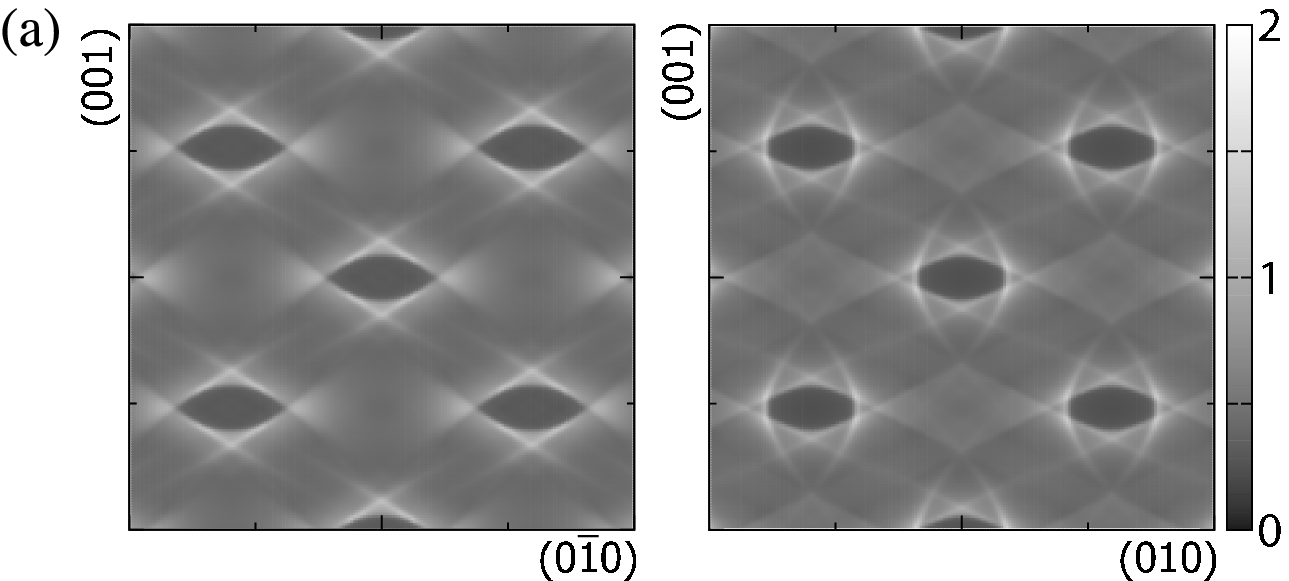}  
\includegraphics[width=7.2cm]{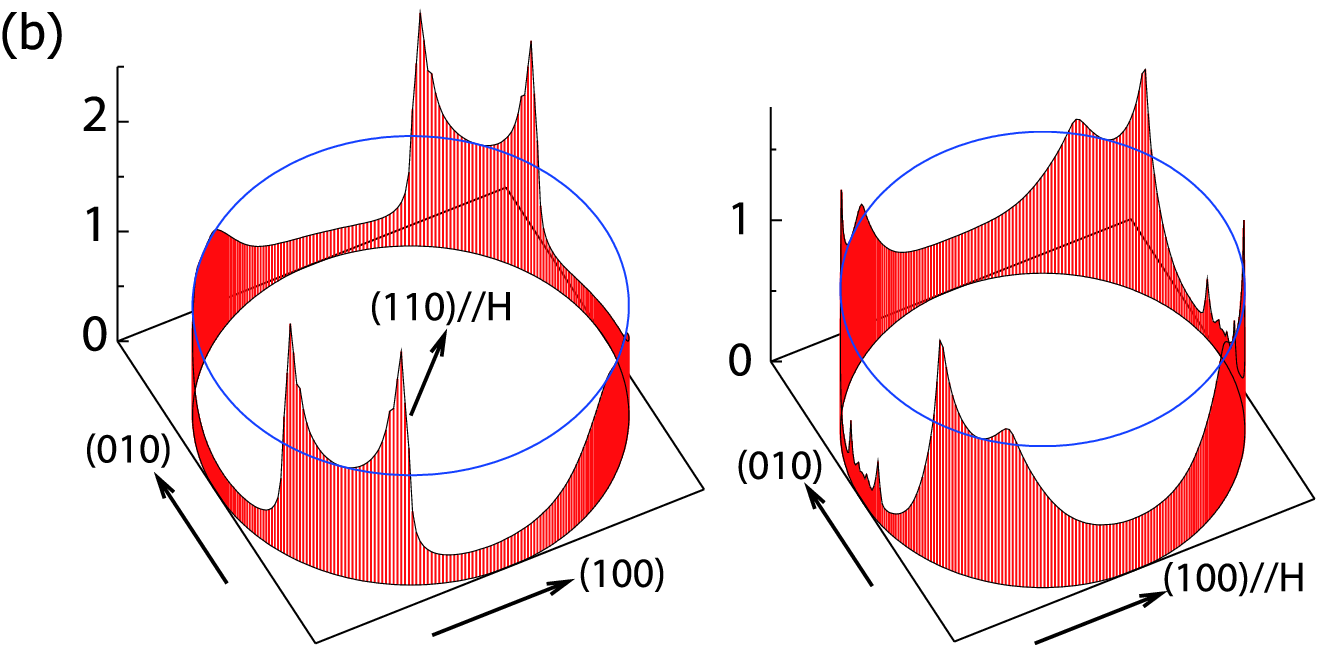}    
\includegraphics[width=7.2cm]{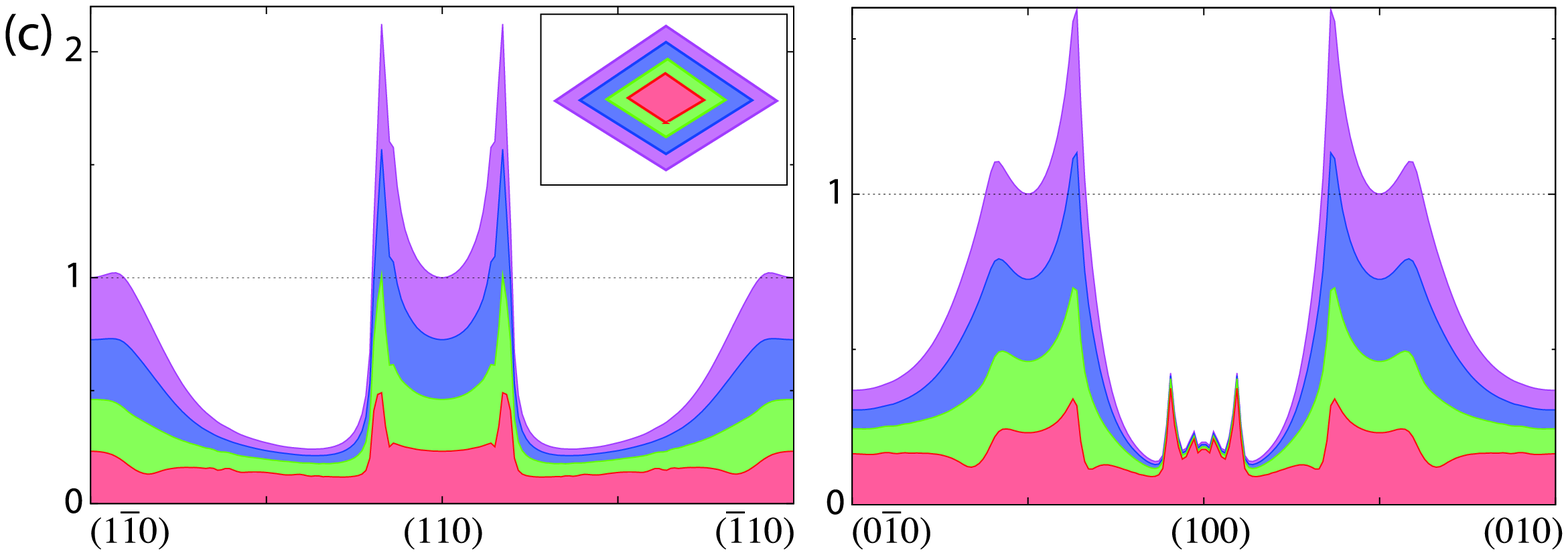}  
\caption{ 
(Color online)
(a) r-DOS $N({\bf r}, E=0)/N_0$ in the real space 
  where vortex cores are placed on the center of each black spot.
  The view ranges are 16$\times$16 in the Eilenberger unit.
  (b) k-DOS $N({\bf k}, E=0)/N_0$ on the Fermi surface.
  Height of the upper rings indicates the normal state value.
  (c) k-DOS is resolved into the contributed real space areas of
  one unit cell shown in the inset. Color codes between them coincide.
  Left column for $H$$\parallel$$(110)$ and right for $H$$\parallel$$(100)$.
  $H/H_{c2}=0.86$, $T/T_c=0.05$ and $\mu=2$.
  }
\label{fig:unit-cell} 
\end{figure} 
%%%%%%%%%%%%%%%%%%%%%%%%%%%%%%%%%%%%%%%% 

The local density of states in real space r-DOS  is given by
$N({\bf r},E)=N_{\uparrow}({\bf r},E)+N_{\downarrow}({\bf r},E)$, where 
$N_\sigma({\bf r},E)
= \langle N_\sigma ({\bf r},{\bf k},E)\rangle_{\bf k}
=N_0 \langle {\rm Re }
\{g( \omega_n +{\rm i} \sigma{\mu} B, {\bf k},{\bf r})
|_{i\omega_n \rightarrow E + i \eta} \}\rangle_{\bf k}$
with $\sigma=1$ ($-1$) for up (down) spin component. 
We typically use $\eta=0.01$.
The total DOS is obtained by the spatial average of the  local DOS as 
$N(E)=N_\uparrow (E) +N_\downarrow (E)
 =\langle N({\bf r},E) \rangle_{\bf r}$. 
 The spectral weight (SW) distribution k-DOS 
 in reciprocal space is given by $N({\bf k},E)=N_{\uparrow}({\bf k},E)+N_{\downarrow}({\bf k},E)$ as
$N_\sigma({\bf k},E)= \langle N_\sigma ({\bf r},{\bf k},E)\rangle_{\bf r}$.
Those r-DOS and k-DOS are complimentary, giving rise to valuable information
for the QP  structures in the vortex state as we will see below.

We show  the r-DOS at $E=0$ in Fig. 1(a) and the corresponding k-DOS
in Figs. 1(b) and (c) for $H$$\parallel$$ (110)$ (left column)  and $H$$\parallel$$ (100)$ (right column)
for the large paramagnetic effect $\mu=2$.
In Fig. 1(a) we see several characteristic and eminent QP trajectories,
notably the bands of the zero energy DOS connecting the nearest neighbor
vortex cores. Those real space trajectories correspond to
the real space motions of the QP induced by $H$.
 It is seen that the zero energy r-DOS is depleted at the vortex core,  
and piles up in the surrounding area because of the Pauli effect.
Due to the Zeeman shift the peak of the zero energy DOS at the vortex core 
moves up and down in the energy, resulting in the depletion of the zero energy DOS
at the core, i.e. the empty core~\cite{ichioka}.
The zero-energy state moves outside from the core where
its maxima occur.
Figure 1(b) shows the corresponding k-DOS where 
on the Fermi circle at $k_z=0$ the SW for each k-direction
is displayed. The SW enhancement along the nodal directions
and strong suppression along the antinodal directions  are clearly seen for 
both field orientations. 
This  comes from the real space area surrounding the core,
which is seen from Fig. 1(c)  where the contribution to the SW
enhancement is resolved in the local areas (see inset) with the same areal size. The core area 
(the central region of the inset) gives
a minor contribution because of the empty core.

%%%%%%%%%%%%%%%%%%%%%%%%%%%%%%%%%%%%%%%% 
\begin{figure}[tb] 
\includegraphics[width=8.0cm]{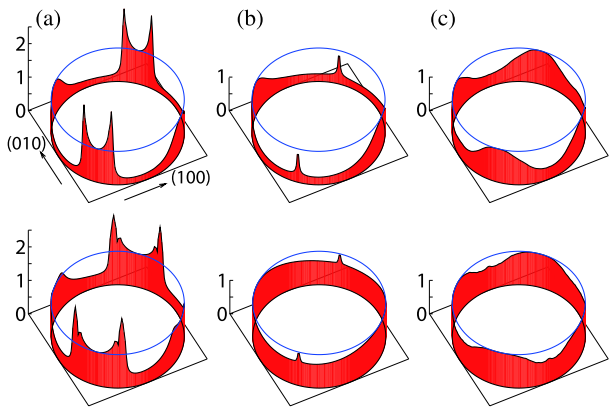} 
 \caption{ 
(Color online)  
k-DOS  $N({\bf k}, E=0)/N_0$ for low field (upper panel) and high
field (lower panel) at $k_z=0$.
(a) For $H$$\parallel$$ (110)$ with Pauli effect $\mu=2.0$ where $H/H_{c2}$=0.74 (upper) and 0.99 (lower).
(b) For the absence of Pauli effect $\mu=0$ where $H/H_{c2}$=0.36 (upper) and 0.86 (lower).
(c)  For $H$$\parallel$$ (001)$ with $\mu=2.0$ where $H/H_{c2}$=0.77 (upper) and 0.98 (lower).
Height of the upper rings indicates the normal state value. 
$T/T_c=0.05$.} 
\label{fig:unit-cell} 
\end{figure} 
%%%%%%%%%%%%%%%%%%%%%%%%%%%%%%%%%%%%%%%% 

The field evolutions of the k-DOS are shown in Fig.2(a)
for $H$$\parallel$$(110)$ with the strong Pauli effect. 
 As $H$ increases towards $H_{c2}$, the k-DOS grows with uneven distribution
 around the Fermi surface, namely the SW is dominated
 along the nodal direction, keeping suppressed along the antinodal direction.
 Upon increasing $H$, the total DOS in the vortex state increases towards
the normal value at $H_{c2}$.
Because of  the Pauli effect, even near $H_{c2}$ the order parameter
amplitude is still non-vanishing. 
Thus beyond a certain $H$ the enhanced SW can
exceed the normal value.

It is contrasted with the case for the absence of the Pauli effect ($\mu=0$)
shown  in  Fig.2(b) where the order parameter decreases continuously
via a second order transition.
 There, the SW never exceeds the normal value and the SW enhancement 
 never occurs even approaching $H_{c2}$.
 Thus the Pauli paramagnetic effect triggers the SDW instability by improving
 nesting condition.
 The SW enhancement occurs because under the in-plane field
 the QP diamagnetic motions are parallel to the $k_z$-axis, thus
 they are sensing the nodal lines running along it,
 giving rise to the singularities in the k-DOS as seen in Fig. 2(a).
 This one-dimensional QP motion along the nodal line never appears
 when $H$$\parallel$$(001)$ as seen below.

 In Fig. 2(c) we display the k-DOS for $H$$\parallel$$(001)$. 
The SW distribution is featureless due to the fact that
the QP trajectories in this field orientation traverses the 
perpendicular nodal line, yielding weaken singularities in the k-DOS.
 There is no SW enhancement above the normal value for any values of $H$,
 implying that the SDW instability is absent in this orientation.
 Thus by tilting away from $H\parallel$(110) towards (001)
 the spectral enhancement ceases to exist.
 According to our calculation, the critical tilting angle
 $\theta_{cr}\sim 30^{\circ}$ from the $ab$ plane.
 Up to this angle the peak intensity of the k-DOS is maintained.
According to Correa, {\it et al}.\cite{correa} the 
 HL phase disappears above $\theta \sim 30^{\circ}$.

%%%%%%%%%%%%%%%%%%%%%%%%%%%%%%%%%%%%%%%% 
\begin{figure}[tb] 
%\vspace{-1cm}

\includegraphics[width=5cm]{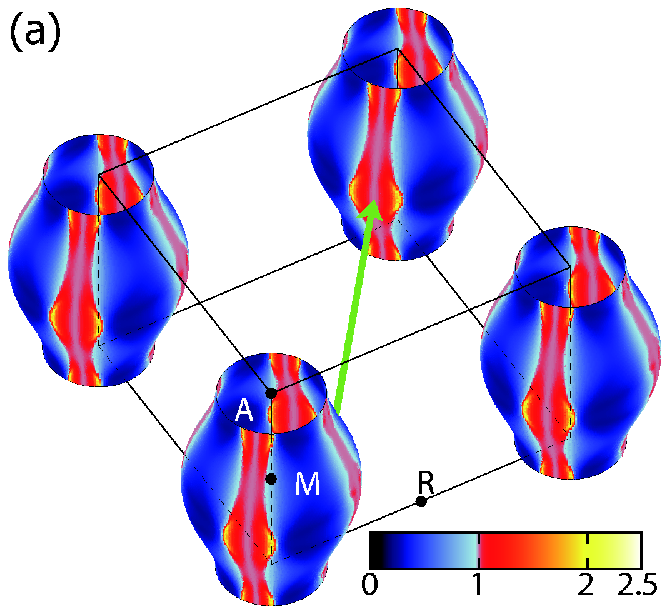}
\begin{minipage}{2.5cm}
\vspace{-4.5cm}
\includegraphics[width=2.5cm]{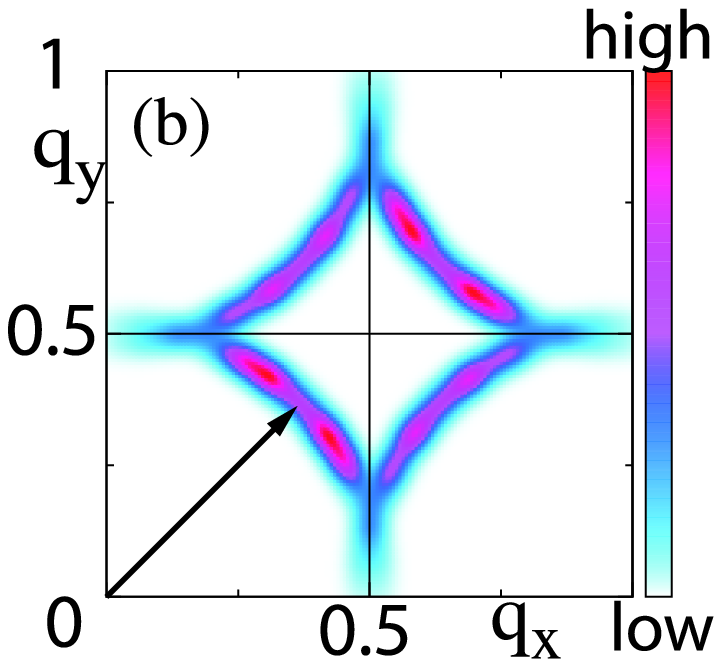}\\
\vspace*{0.5\baselineskip}
\includegraphics[width=2.0cm]{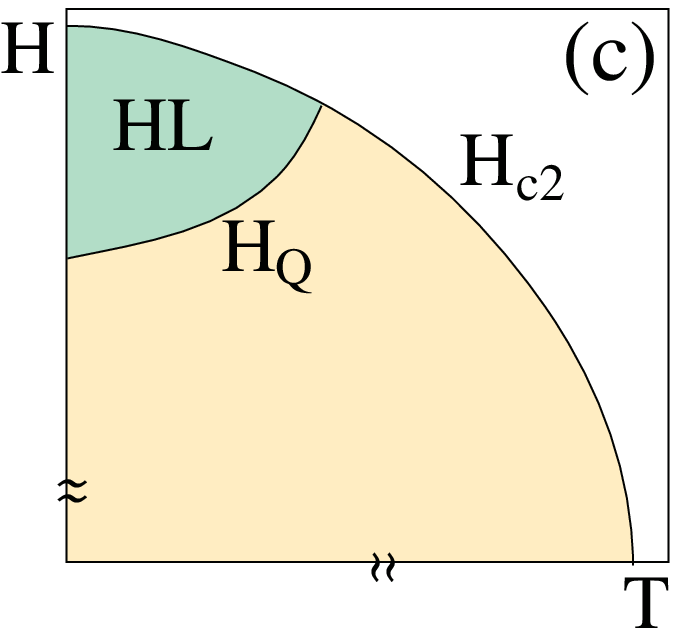}  
\end{minipage}

\caption{ 
(Color online)  
(a) Spectral weight distribution drown on the Fermi surfaces
situated on the four corners of the Brillouin zone.
$H/H_{c2}=0.86$, $T/T_c=0.05$, $H\parallel (110)$ and $\mu=2$.
Arrow corresponds to $q=0.05$ when $k_F=0.26$. 
(b) Corresponding joint DOS $N_J({\bf q})$.
Arrow  denotes the best nesting vector ${\bf Q}$
in $0<q_x,q_y<1.0$ and $q_z=0.5$, indicating the four
maxima ${\bf Q}=(0.5\pm q, 0.5\pm q, 0.5)$
centered around $(0.5,0.5,0.5)$. 
(c) Schematic phase diagram of the HL phase.
} 
\label{fig:unit-cell} 
\end{figure} 
%%%%%%%%%%%%%%%%%%%%%%%%%%%%%%%%%%%%%%%% 

  In Fig. 3(a) we display the 3D views of the SW distributions on the $\alpha$-Fermi surfaces in
 CeCoIn$_5$. This Fermi surface is situated in the four corner of
 the Brillouin zone. It is clear that the best  nesting is expected for ${\bf Q}=(q,q,0.5)$
 in reciprocal vector units, where $q$ is evaluated geometrically as $q=1-2k_F$ as indicated by arrows.
 The $k_z$ component of ${\bf Q}$ comes from the warping of the $\alpha$-Fermi surface
 along the $k_z$ direction.
 In fact, the joint density of states (JDOS) defined by 
 $N_J({\bf q})=\langle N({\bf k}, E=0)N({\bf q}-{\bf k}, E=0)\rangle_{\bf k}$,  is one of 
the indicators to check the degree of the nesting condition and whose maximum
may signify the SDW instability. 
The JDOS is presented in Fig. 3(b).
 The optimal nesting vector appears at $Q=(0.5\pm q,0.5\pm q,0.5)$
 with $q=0.05$ when we approximate  the Fermi wave number as $k_F=0.26$ of the $\alpha$-orbit
at the $k_z=0$ plane  according to the combined efforts by the dHvA experiment
and band calculation\cite{harima}.
It is not apparent whether the SDW is described by the double $Q$
or the single $Q$. Theoretically those are equally possible.
Experimentally it is desirable to determine it.
In particular, for $H$$\parallel$$(100)$ those pairs of the nesting vectors are
equivalent by symmetry, leading to either double $Q$ structure or
single $Q$ structure with two domains. 
Yet, the existing experiment\cite{michel2} only observes one of the pair
 ${\bf Q}=(0.5-q,0.5-q,0.5)$ and  ${\bf Q}=(0.5+q,0.5+q,0.5)$.

An SDW instability with the nesting vector ${\bf Q}$ could occur
beyond a certain high field $H_Q(T)$ and below $H_{c2}$ (see Fig. 3(c) for schematic phase diagram).
As $H$ increases towards $H_{c2}$, the SW enhancement by vortices becomes larger only along the nodal direction
and provides the best SDW nesting condition.
It occurs through the repulsive interaction
$U$ between those induced QPs. The Stoner instability condition; $U\chi({\bf Q})>1$
with $\chi({\bf Q})$ static susceptibility of ${\bf Q}$ component is fulfilled.
Note that $\chi({\bf Q}=0)=\mu_B^2N_0$.
Since the observed wave length of the SDW modulation is an order of $\sim 5$nm
and much shorter than $\xi\sim 30$nm, there is still a possibility for FFLO to occur
as advocated by several authors\cite{littlewood,tsunetsugu,yanase,miyake,ikeda}
because the present SDW polarized along the $c$-axis \cite{michel1,michel2} does not liberate the penalty due to the Zeeman effect
longitudinal to the in-plane fields.
The in-plane anisotropy of the transition line $H_Q(T)$ may not be large
because the total DOSs for two directions (100) and (110) differ only by at most
0.4$\%$\cite{an}.
As increasing $T$ under a fixed $H$, the peak value of the SW enhancement $N(k, E=0)$ decreases
because of thermal broadening of the singularities of k-DOS. Thus
$dH_Q(T)/dT>0$, coinciding with the observation\cite{losalamos}.
Likewise impurities easily kill the singularities, thus doping experiment
can be understood where the HL phase area diminishes upon small Cd- and Hg-dopings\cite{tokiwa}.
As shown above, the tilting field towards the $c$-axis ceases to  
induce the SW enhancement above the normal value.
This is supported by  the recent neutron experiment\cite{blackburn} and by Correa, {\it et al}.\cite{correa}.
Therefore the  HL phase in the in-plane field is not connected to the HL phase in the $c$-axis field,
implying that the latter cannot be an SDW instability, but may be genuine FFLO.
Those are logical conclusions and predictions based on our microscopic calculations.

In summary, by solving the microscopic Eilenberger equations self-consistently
with the $d_{x^2-y^2}$ pairing state and strong Pauli paramagnetic effect,
we obtain the detailed information on the quasiparticle structures in both real
and reciprocal spaces for vortex lattice states.
This complemental information allows us to draw a picture 
that for the in-plane fields the spectral weight is enhanced by vortices
for the nodal direction in k-space exceeded above the normal value,
and thus signals an SDW instability whose ordering vector directed
towards the nodal direction.
This kind of the coexistence between SDW and SC with vortices is a novel form
confined exclusively inside SC (see Fig.3(c)), which is beyond known
coexistence scheme where two orderings compete each other
to try to open their own energy gaps on the Fermi surface\cite{machida}.
The proposed new SDW mechanism may apply other heavy Fermion superconductors,
such as Ce$_2$PdIn$_8$\cite{dong}.

%%%%%%%%%%%%%%%%%%%%%%%%%%%%%%%%%%%%%%%%%%%%%%%%%%%%%%%%%%%%%%%% 
%\section*{Acknowledgments}  
  
The authors are grateful for insightful discussions with  
M. Kenzelmann, S. Gerber, J. S. White, J. L. Gavilano, T. Sakakibara, E. M. Forgan,
K. Kumagai and R. Ikeda.

%%%%%%%%%%%%%%%%%%%%%%%%%%%%%%%%%%%%%%%%%%%%%%%%%%%%%%%%%%%%%%%% 


\begin{thebibliography}{99} 
 
 \bibitem{chris}  
C. Pfleiderer, Rev. Mod. Phys. {\bf 81}, 1551 (2009). 

 \bibitem{mesot}  
 B. Lake, {\it et al}., Nature  {\bf 415}, 299  (2002). 
 J. Chang, {\it et al}., Phys. Rev. Lett. {\bf 102}, 177006  (2009). 
 
  \bibitem{flouquet}
  G. Knebel, {\it et al}., arXiv: 0911.5223.
  
  
 \bibitem{abrikosov}
 A. A. Abrikosov, Soviet Phys-JETP  {\bf 5}, 1174  (1957). 
 
 
  
 \bibitem{hasegawa}
 M. Ichioka, {\it et al}., Phys. Rev. B {\bf 59}, 184 and 8902 (1999). 

   
 
 \bibitem{hess}
 H. F. Hess, {\it et al}., Phys. Rev. Lett. {\bf 62}, 214 (1989). 
 N. Nakai, {\it et al}., {\it ibid} {\bf 97}, 147001 (2006). 


 \bibitem{bianchi}
A. D. Bianchi, {\it et al}., Science {\bf 319}, 177 (2008).


 \bibitem{an}
 K. An, {\it et al}., Phys. Rev. Lett. {\bf 104}, 037002 (2010). 


 \bibitem{izawa}
K. Izawa, {\it et al}., Phys. Rev. Lett. {\bf 87}, 057002 (2001). 
 
 
  \bibitem{littlewood}
  A. Aperis, {\it et al}., Phys. Rev. Lett. {\bf 104}, 216403 (2010). 
  
 \bibitem{tsunetsugu}
 D. F. Agterberg, {\it et al}., Phys. Rev. Lett. {\bf 102}, 207004 (2009). 
 

 \bibitem{yanase}
 Y. Yanase and M. Sigrist,  J. Phys. Soc. Jpn. {\bf 78}, 114715 (2009). 

 
 \bibitem{miyake}
 K. Miyake, J. Phys. Soc. Jpn. {\bf 77}, 123703 (2008). 
 
 \bibitem{ikeda}
 R. Ikeda, {\it et al}., Phys. Rev. B{\bf 82}, 060510(R) (2010). 
 
 
 \bibitem{losalamos}
 A. D. Bianchi, {\it et al}., Phys. Rev. Lett. {\bf 91}, 187004 (2003). 
 
 \bibitem{kakuyanagi}
 K. Kakuyanagi, {\it et al}., Phys. Rev. Lett. {\bf 94}, 047602 (2005). 
 
 \bibitem{vesna}
 G. Koutroulakis, {\it et al}., Phys. Rev. Lett. {\bf 101}, 047004 (2008). 
 
 \bibitem{curro}
 B. -L. Young, {\it et al}., Phys. Rev. Lett. {\bf 98}, 036402 (2007). 
 
 \bibitem{matsuda}
 T. Watanabe, {\it et al}., Phys. Rev. B {\bf 70}, 020506(R) (2004). 
 
 \bibitem{kumagai}
 K. Kumagai, {\it et al}., Phys. Rev. Lett. {\bf 97}, 227002 (2006). 

\bibitem{michel1}
 M. Kenzelmann, {\it et al}., Science {\bf 321}, 1652 (2008).
 
 \bibitem{michel2}
 M. Kenzelmann, {\it et al}., Phys. Rev. Lett. {\bf 104}, 127001 (2010). 
 
  
\bibitem{aoki}
H. Aoki {\it et al}., J. Phys. Condens. Matter {\bf 16}, L13 (2004).

 \bibitem{ichioka}
 M. Ichioka and K. Machida, Phys. Rev. B {\bf 76}, 064502 (2007). 

  \bibitem{harima}
  H. Shishido, {\it et al}., J. Phys. Soc. Jpn. {\bf 71}, 162 (2002). 

 
  \bibitem{correa}
V. F. Correa, {\it et al}., Phys. Rev. Lett. {\bf 98}, 087001 (2007). 


 \bibitem{tokiwa}
Y. Tokiwa, {\it et al}., Phys. Rev. Lett. {\bf 101}, 037001 (2008). 

 \bibitem{blackburn}
E. Blackburn, {\it et al}., preprint.


\bibitem{machida}
K. Machida and M. Kato, Phys. Rev. Lett. {\bf 58}, 1986 (1987). 
K. Machida, J. Phys. Soc. Jpn. {\bf 50}, 2195 (1981). 
K. Machida and T. Matsubara, {\it ibid} {\bf 50}, 3231 (1981). 

\bibitem{dong}
J. K. Dong, {\it et al}, arXiv:1008.0679.

\end{thebibliography}
\end{document}